\documentclass[prb,twocolumn,floatfix,superscriptaddress]{revtex4}

\usepackage{graphicx}
\usepackage{amsmath}
\usepackage{amssymb}
\usepackage[colorlinks=true,citecolor=blue,linkcolor=blue]{hyperref}
\usepackage{amsfonts}
\usepackage{color,xcolor}
\usepackage{epstopdf}
\usepackage{braket}
\usepackage{bm}
\usepackage{bbold}

\renewcommand{\vec}[1]{\ensuremath{\boldsymbol{#1}}}

\begin{document}

\title{Interlayer excitons in transition metal dichalcogenide heterostructures}
\date{\today}
\author{M. Van der Donck}
\email{matthias.vanderdonck@uantwerpen.be}
\affiliation{Department of Physics, University of Antwerp, Groenenborgerlaan 171, B-2020 Antwerp, Belgium}
\author{F. M. Peeters}
\email{francois.peeters@uantwerpen.be}
\affiliation{Department of Physics, University of Antwerp, Groenenborgerlaan 171, B-2020 Antwerp, Belgium}

\begin{abstract}
Starting from the single-particle Dirac Hamiltonian for charge carriers in monolayer transition metal dichalcogenides (TMDs), we construct a four-band Hamiltonian describing interlayer excitons consisting of an electron in one TMD layer and a hole in the other TMD layer. An expression for the electron-hole interaction potential is derived, taking into account the effect of the dielectric environment above, below, and between the two TMD layers as well as polarization effects in the transition metal layer and in the chalcogen layers of the TMD layers. We calculate the interlayer exciton binding energy and average in-plane interparticle distance for different TMD heterostructures. The effect of different dielectric environments on the exciton binding energy is investigated and a remarkable dependence on the dielectric constant of the barrier between the two layers is found, resulting from competing effects as a function of the in-plane and out-of-plane dielectric constants of the barrier. The polarization effects in the chalcogen layers, which in general reduce the exciton binding energy, can lead to an increase in binding energy in the presence of strong substrate effects by screening the substrate. The excitonic absorbance spectrum is calculated and we show that the interlayer exciton peak depends linearly on a perpendicular electric field, which agrees with recent experimental results.
\end{abstract}

\maketitle

\section{Introduction}

Monolayer transition metal dichalcogenides (TMDs) such as MoS$_2$, MoSe$_2$, WS$_2$, WSe$_2$, etc.\cite{mak1,splendiani,mak2,zeng,cao,ross}, lack inversion symmetry, which leads to a direct band gap at the corners of the hexagonal first Brillouin zone. This allows for the optical excitation of excitonic states\cite{korn,sallen,mak3,he,chernikov}, i.e. bound systems of an electron and a hole. Monolayer TMDs are strictly two dimensional (2D) systems and as a result the excitons in these systems are very tightly bound, i.e. they have binding energies of the order of several hundreds of meV, which is two orders of magnitude larger as compared to excitons in conventional three dimensional semiconductors\cite{elliot,kulak,expT0,francoisE,francoisT}.

More recently attention has turned towards assembling and studying van der Waals heterostructures\cite{vdwstructure}. This includes the possibility of stacking different kinds of TMDs on top of each other. When this stacking results in a type-II band alignment, which is predicted to occur in a wide range of TMD heterostructures\cite{type1,type2,type3,typeII,type4}, it is possible to optically excite so-called interlayer excitons. These are excitons consisting of an electron localized in one of the TMD layers and a hole localized in the other TMD layer and play a crucial role in excitonic superfluidity\cite{super1,super2,super3,super4,decouple4}. These interlayer excitons, which were detected in recent experiments\cite{exp1,exp2,exp3,exp4,exp5}, have a binding energy, of the order of hundred meV, and their lifetime is one to two orders of magnitude larger than that of intralayer excitons\cite{exp2,exp3}.

There are a few theoretical works studying interlayer excitons in TMD heterostructures, for example by using ab initio many-body perturbation theory with the Bethe-Salpeter equation\cite{vacuum}, which is computationally demanding, and by solving the effective mass Wannier equation\cite{sio2}. In the present paper we construct a theoretical two-body massive Dirac model for describing interlayer excitons, including a complete description of the interlayer electron-hole interactions obtained by solving the Poisson equation for a general heterostructure system, which allows to calculate exciton binding energies, in-plane interparticle distances, and the excitonic absorbance spectrum. This model has the advantages of being computationally fast and yet still allowing to include the effects of the valence bands. We use it to investigate the effect of different dielectric environments and polarization effects in the chalcogen layers on the interlayer exciton binding energy and we compare our calculated excitonic absorbance spectrum and electric field dependent interlayer exciton energies to recent experimental results.

Our paper is organized as follows. In Sec. \ref{sec:Model} we present an outline of the exciton model and the calculation of the interlayer electron-hole interaction potential. The numerical results are discussed in Sec. \ref{sec:Results}. We summarize the main conclusions in Sec. \ref{sec:Summary and conclusion}.

\section{Model}
\label{sec:Model}

\subsection{Exciton Hamiltonian}

We start from the effective low-energy single-electron Hamiltonian\cite{theory1} in the basis $\mathcal{B}^e_{s,\tau}=\{\ket{\phi^e_{c,s,\tau}},\ket{\phi^e_{v,s,\tau}}\}$ spanning the 2D Hilbert space $\mathcal{H}^e_{s,\tau}$, with $\ket{\phi^e_{c,s,\tau}}$ and $\ket{\phi^e_{v,s,\tau}}$ the atomic orbital states at the conduction $(c)$ and valence $(v)$ band edge, respectively:
\begin{equation}
\label{singelhame}
\begin{split}
H^e_{s,\tau}(\vec{k}) = at(\tau k_x\sigma_x+k_y\sigma_y)+\frac{\Delta}{2}\sigma_z+\frac{\lambda s\tau}{2}(I_2-\sigma_z),
\end{split}
\end{equation}
where $\sigma_i$ ($i=x,y,z$) are Pauli matrices, $I_2$ is the two by two identity matrix, $a$ the lattice constant, $t$ the hopping parameter, $\tau=\pm1$ the valley index, $s=\pm1$ the spin index, $\Delta$ the band gap, and $\lambda$ the spin-orbit coupling strength leading to a spin splitting of $2\lambda$ at the valence band edge.

Since a hole with wave vector $\vec{k}$, spin $s$, and valley index $\tau$ is the absence of an electron with opposite wave vector, spin, and valley index, the single-hole Hamiltonian can immediately be obtained from the single-electron Hamiltonian as $\hat{H}^h_{s,\tau}(\vec{k})=-\hat{H}^e_{-s,-\tau}(-\vec{k})$, and the eigenstates of this Hamiltonian span the Hilbert space $\mathcal{H}^h_{s,\tau}$. The total exciton Hamiltonian acts on the product Hilbert space spanned by the tensor products of the single-particle states at the band edges, $\mathcal{B}^{exc}_{\alpha}=\mathcal{B}^e_{s^e,\tau^e}\otimes\mathcal{B}^h_{s^h,\tau^h}$, and is given by
\begin{equation}
\label{excham}
\begin{split}
H^{exc}_{\alpha}(\vec{k}^e,\vec{k}^h,r_{eh},h) = &H^e_{s^e,\tau^e}(\vec{k}^e)\otimes I_2 \\
&\hspace{-10pt}-I_2\otimes H^e_{-s^h,-\tau^h}(-\vec{k}^h)-V(r_{eh},h)I_{4},
\end{split}
\end{equation}
where $\alpha$ is a shorthand notation for $\{s^e,\tau^e,s^h,\tau^h\}$ and the electron-hole interaction potential $V(r_{eh},h)$ is derived in the next subsection with $r_{eh}=|\vec{r}_e-\vec{r}_h|$ the in-plane distance between the electron and the hole and $h$ the interlayer distance. Explicitly writing out the different matrix elements gives
\begin{widetext}
\begin{equation}
\label{hamtot}
H^{exc}_{\alpha}(\vec{k}^e,\vec{k}^h,r_{eh},h) =
\begin{pmatrix}
\delta_1-V(r_{eh},h) & a^ht^h(-\tau^hk_x^h-ik_y^h) & a^et^e(\tau^ek_x^e-ik_y^e) & 0 \\
a^ht^h(-\tau^hk_x^h+ik_y^h) & \delta_2-V(r_{eh},h) & 0 & a^et^e(\tau^ek_x^e-ik_y^e) \\
a^et^e(\tau^ek_x^e+ik_y^e) & 0 & \delta_3-V(r_{eh},h) & a^ht^h(-\tau^hk_x^h-ik_y^h) \\
0 & a^et^e(\tau^ek_x^e+ik_y^e) & a^ht^h(-\tau^hk_x^h+ik_y^h) & \delta_4-V(r_{eh},h)
\end{pmatrix},
\end{equation}
\end{widetext}
with
\begin{equation}
\label{diagonaalelem}
\begin{split}
&\delta_1 = \frac{\Delta^e-\Delta^h}{2}, \\
&\delta_2 = \frac{\Delta^e+\Delta^h}{2}-\lambda^hs^h\tau^h, \\
&\delta_3 = -\frac{\Delta^e+\Delta^h}{2}+\lambda^es^e\tau^e, \\
&\delta_4 = -\frac{\Delta^e-\Delta^h}{2}+\lambda^es^e\tau^e-\lambda^hs^h\tau^h.
\end{split}
\end{equation}
The eigenvalue problem for this Hamiltonian,
\begin{equation}
\label{eigen}
H^{exc}_{\alpha}(\vec{k}^e,\vec{k}^h,r_{eh},h)\ket{\Psi^{exc}_{\alpha}} = E^{exc}_{\alpha}(\vec{k}^e,\vec{k}^h)\ket{\Psi^{exc}_{\alpha}},
\end{equation}
defines the interlayer exciton energy $E^{exc}_{\alpha}(\vec{k}^e,\vec{k}^h)$ and the interlayer exciton eigenstate $\ket{\Psi^{exc}_{\alpha}}=\left(\ket{\phi^{e,h}_{c,c}},\ket{\phi^{e,h}_{c,v}},\ket{\phi^{e,h}_{v,c}},\ket{\phi^{e,h}_{v,v}}\right)^T$, where the subscript $\alpha$ and the superscript $exc$ have been dropped in the right hand side for notational clarity. In this work we will always consider $\alpha=\{1,1,-1,-1\}$. The above eigenvalue problem is a matrix equation which can, following a procedure analogous to earlier works \cite{decouple4,decouple1,decouple2,decouple3,previous}, be decoupled into a single equation. Transforming to center of mass and relative coordinates, taking the center of mass momentum to be zero, and assuming the electron and hole kinetic energies to be small compared to the band gap and the exciton energy, this equation reduces for $s$-state excitons to
\begin{widetext}
\begin{equation}
\label{difvgl}
\begin{split}
\bigg[&-\left(a^ht^h\right)^2\left(\frac{1}{E^{exc}_{\alpha}+V(r,h)-\delta_1}\nabla^2_{\vec{r}}+\frac{\partial}{\partial r}\left(\frac{1}{E^{exc}_{\alpha}+V(r,h)-\delta_1}\right)\frac{\partial}{\partial r}\right)+\delta_2-V(r,h) \\
&-\left(a^et^e\right)^2\left(\frac{1}{E^{exc}_{\alpha}+V(r,h)-\delta_4}\nabla^2_{\vec{r}}+\frac{\partial}{\partial r}\left(\frac{1}{E^{exc}_{\alpha}+V(r,h)-\delta_4}\right)\frac{\partial}{\partial r}\right]\phi^{e,h}_{c,v}(r) = E^{exc}_{\alpha}\phi^{e,h}_{c,v}(r),
\end{split}
\end{equation}
\end{widetext}
where $\phi^{e,h}_{c,v}(r)$ is the component of the exciton eigenstate representing an exciton consisting of an electron in the conduction band and a hole in the valence band. Note that our choice of $\alpha$ and the zero center of mass momentum imply that we study optically active excitons. We presented a detailed derivation of this equation and the expressions for calculating the other three components of the exciton eigenstate in Appendix A of Ref. [\onlinecite{previous}]. The above equation is a differential eigenvalue equation, which we solve with the finite element method, with the additional complication of the eigenvalue appearing in the left hand side as well. Therefore we have to solve this equation self-consistently by choosing an initial value for $E^{exc}_{\alpha}$ and inserting it in the left hand side and numerically calculating the corresponding eigenvalue in the right hand side. This newly calculated eigenvalue is subsequently used in the left hand side to calculate a new eigenvalue. This is repeated until convergence is reached. After the exciton energy $E^{exc}_{\alpha}$ is obtained, the binding energy is obtained from
\begin{equation}
\label{bindingenergy}
E_{b,\alpha}^{exc}=\frac{\Delta^e+\Delta^h}{2}-\lambda^hs^h\tau^h-E^{exc}_{\alpha}.
\end{equation}

\subsection{Electron-hole interaction}

\begin{figure}
\centering
\includegraphics[width=8.5cm]{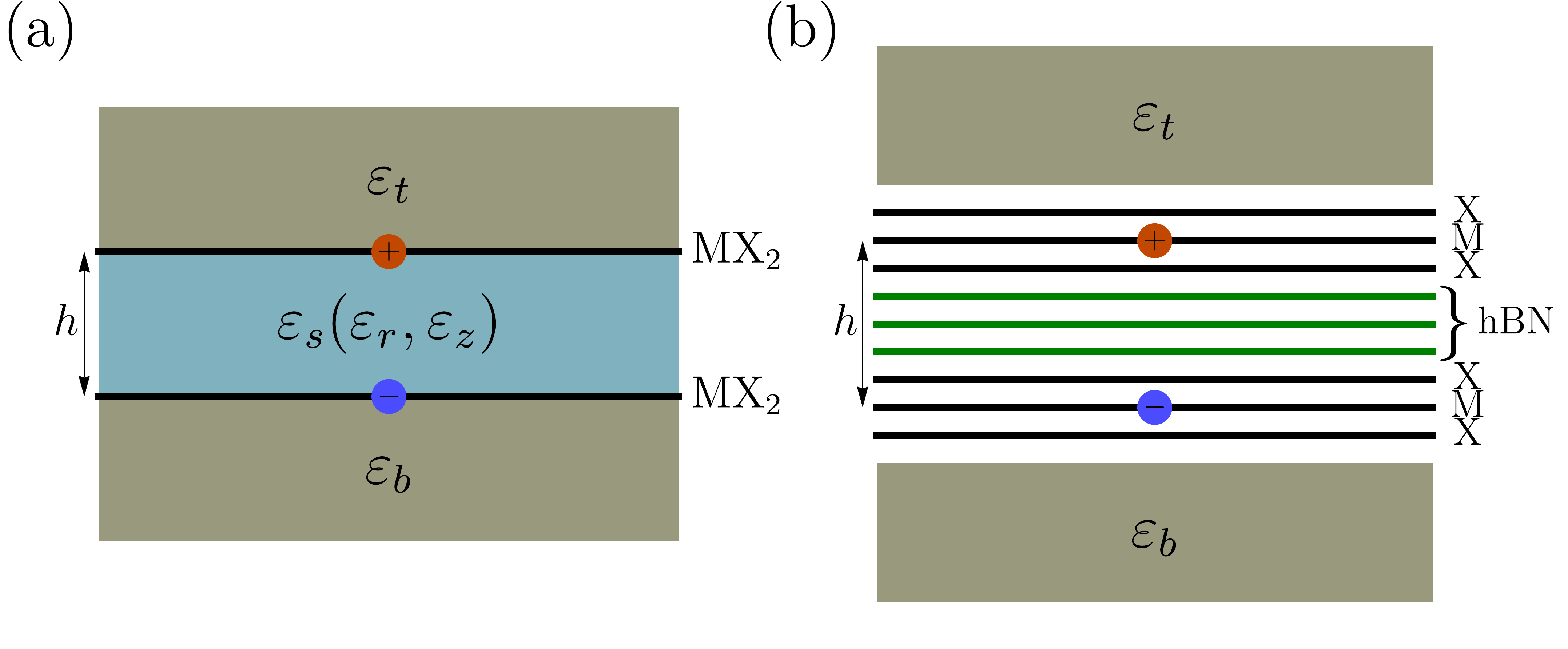}
\caption{(Color online) Schematic of a TMD heterostructure with substrates with isotropic dielectric constants $\varepsilon_b$ and $\varepsilon_t$ below and above the system, respectively. (a) The TMDs are modeled by monolayers with 2D polarizabilities $\chi_{2\text{D}}^b$ and $\chi_{2\text{D}}^t$ for the bottom and top TMD layer, respectively. The barrier between the two layers has a dielectric constant $\varepsilon_r$ and $\varepsilon_z$ parallel and perpendicular to the plane, respectively. (b) The TMDs are modeled by trilayers with 2D polarizabilities $\chi_{2\text{D}}^{i,M}$ and $\chi_{2\text{D}}^{i,X}$ for the transition metal and chalcogen layers, respectively, with $i=b,t$. In between the two TMDs are hBN layers with 2D polarizability $\chi_{2\text{D}}^{hBN}$.}
\label{fig:schema}
\end{figure}

Excitons in a single TMD layer are governed by the intralayer interaction potential which, due to non-local screening effects, is given by\cite{screening1,screening2,screening3}
\begin{equation}
\label{interpot}
V^{intra}(r_{ij}) = \frac{e^2}{4\pi\kappa\varepsilon_0}\frac{\pi}{2r_0}\left[H_0\left(\frac{r_{ij}}{r_0}\right)-Y_0\left(\frac{r_{ij}}{r_0}\right)\right],
\end{equation}
with $r_{ij}=|\vec{r}_i-\vec{r}_j|$, where $Y_0$ and $H_0$ are, respectively, the Bessel function of the second kind and the Struve function, with $\kappa=(\varepsilon_b+\varepsilon_t)/2$ where $\varepsilon_{b(t)}$ is the dielectric constant of the environment below (above) the TMD monolayer, and with $r_0=\chi_{2\text{D}}/(2\kappa)$ the screening length where $\chi_{2\text{D}}$ is the 2D polarizability of the TMD. For $r_0=0$ this potential reduces to the bare Coulomb potential $V(r_{ij})=e^2/(4\pi\kappa\varepsilon_0r_{ij})$. Increasing the screening length leads to a decrease in the short-range interaction strength while the long-range interaction strength is unaffected. For very large screening lengths $r_0\rightarrow\infty$ the interaction potential becomes logarithmic, i.e. $V(r_{ij})=e^2/(4\pi\kappa\varepsilon_0r_0)\text{ln}(r_0/r_{ij})$.

The electron-hole interaction which binds the interlayer exciton differs considerably from the above intralayer interactions. An expression for this interlayer interaction potential can be found by starting from Gauss's law and is derived in Appendix \ref{sec:appA}.
\begin{figure}
\centering
\includegraphics[width=8.5cm]{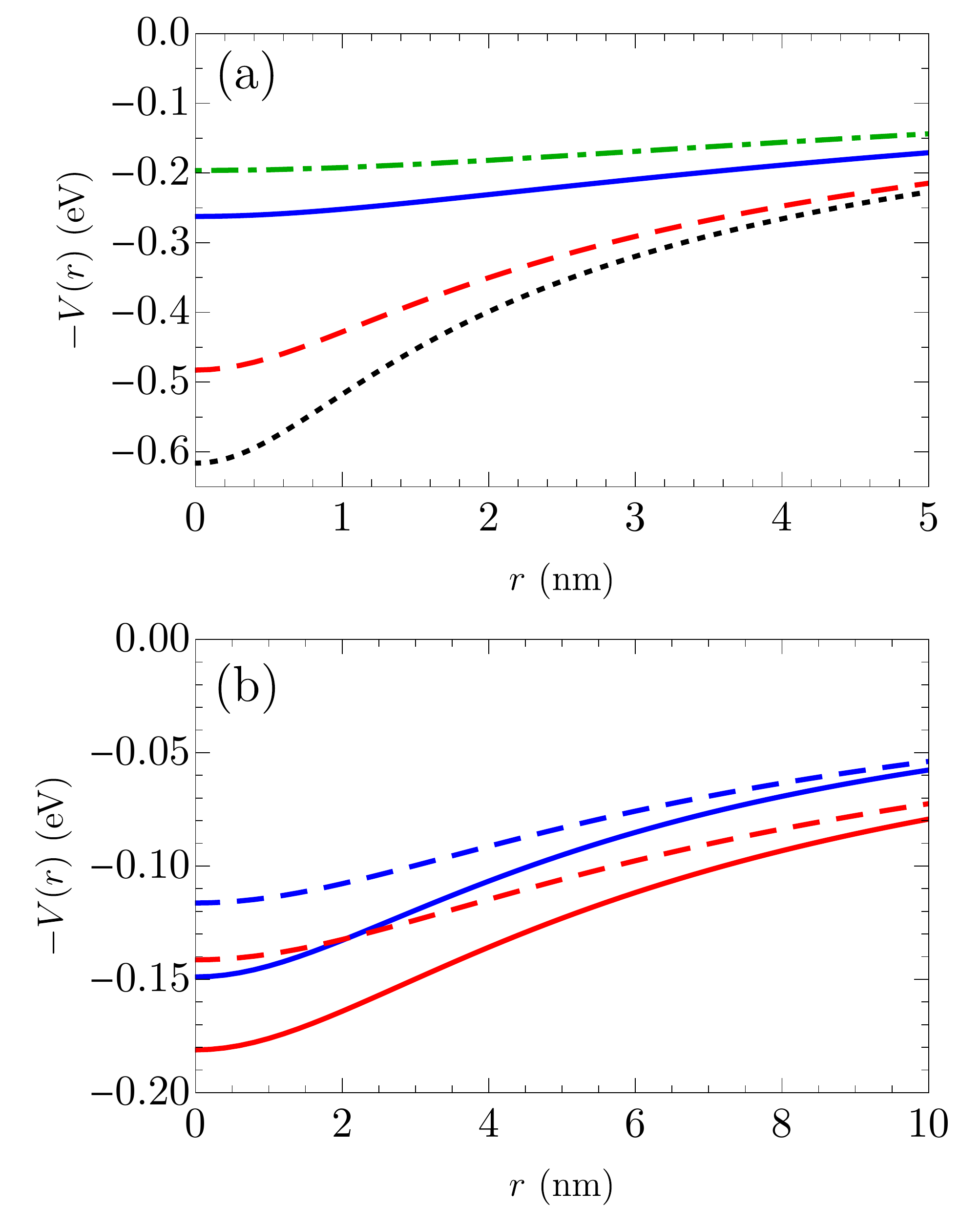}
\caption{(Color online) (a) Solid, blue: Interlayer interaction potential between a hole and an electron in a TMD heterostructure for $\varepsilon_b=\varepsilon_t=\varepsilon_s=1$, interlayer distance $h=1$ nm, and with 2D polarizabilities $\chi_{2\text{D}}^b=\chi_{2\text{D}}^t=8$ nm. Dashed, red: Same as the previous but now for $\chi_{2\text{D}}^t=0$. Dot-dashed, green: Interlayer interaction potential taking into account the polarizabilities of the chalcogen layers with $\chi_{2\text{D}}^{b,X}=\chi_{2\text{D}}^{t,X}=2$ nm. Dotted, black: Intralayer interaction potential with $r\rightarrow\sqrt{r^2+h^2}$. (b) Interlayer interaction potential for a MoS$_2$-WS$_2$ heterostructure on SiO$_2$ (blue) and between hBN layers (red) and with 1 layer of hBN between the two TMDs which are modeled by trilayers with (dashed) and without (solid) polarization effects in the chalcogen layers with $\chi_{2\text{D}}^X=\chi_{2\text{D}}^M/4$. We use $\varepsilon_b=3.8$ for SiO$_2$ and ($\varepsilon_r=4.5,\varepsilon_z=1$) for the hBN substrates. For the hBN layer in the barrier we use $\chi_{2\text{D}}^{\text{hBN}}=1.17$ nm. For Mo and for W we use $\chi_{2\text{D}}^{\text{Mo}}=8.29$ nm and $\chi_{2\text{D}}^{\text{W}}=7.58$ nm, respectively.}
\label{fig:interactie}
\end{figure}
In Fig. \ref{fig:schema} we show two possible models of a TMD heterostructure. In (a) the TMDs are modeled by monolayers and the barrier between the two TMDs is modeled by a 3D homogeneous material. The substrates above ($t$) and below ($b$) the TMD heterostructure are assumed to be isotropic, i.e. $\varepsilon_r^{t/b}=\varepsilon_z^{t/b}=\varepsilon_{t/b}$, whereas for the barrier between the two TMDs the more general case in which $\varepsilon_r$ and $\varepsilon_z$ can differ from each other is considered. For the interaction between a charge in the top layer ($z'=h/2$) and a charge in the bottom layer ($z=-h/2$) we find the following wave vector dependent dielectric function:
\begin{equation}
\label{epsa}
\begin{split}
&\varepsilon\left(q,-\frac{h}{2},\frac{h}{2}\right) = \frac{\varepsilon_b+\varepsilon_t+(\chi_{2\text{D}}^b+\chi_{2\text{D}}^t)q}{2}\cosh\left(\sqrt{\frac{\varepsilon_r}{\varepsilon_z}}hq\right) \\
&\hspace{30pt}+\frac{\varepsilon_r\varepsilon_z+(\varepsilon_b+\chi_{2\text{D}}^bq)(\varepsilon_t+\chi_{2\text{D}}^tq)}{2\sqrt{\varepsilon_r\varepsilon_z}}\sinh\left(\sqrt{\frac{\varepsilon_r}{\varepsilon_z}}hq\right),
\end{split}
\end{equation}
with $h$ the distance between the two TMDs and with $\chi_{2\text{D}}^b$ ($\chi_{2\text{D}}^t$) the 2D polarizability of the bottom (top) TMD. In general, no analytic expression can be found for the real space interaction potential \eqref{interactiereeel} and one has to resort to numerical integration. The results are shown in Fig. \ref{fig:interactie}(a). This shows that the interaction potential is considerably weaker than what is found by simply substituting $r\rightarrow\sqrt{r^2+h^2}$ in the intralayer interaction potential \eqref{interpot}. The only limits for which an analytic expression for the real space interaction potential can be found are: 1) $h=0$, for which $\varepsilon(q)=\kappa+\chi_{\text{2D}}q/2$ and the interaction potential reduces to the intralayer potential \eqref{interpot} (when one of the 2D polarizabilities is set to 0) and 2) $\chi_{2\text{D}}^b=\chi_{2\text{D}}^t=0$ and $\varepsilon_b=\varepsilon_t=\varepsilon_r=\varepsilon_z=1$, for which $\varepsilon(q,h)=e^{hq}$ and the interaction potential reduces to $q_1q_2/(4\pi\varepsilon_0\sqrt{r^2+h^2})$.

In Fig. \ref{fig:schema}(b) the TMDs are modeled by trilayers, i.e. a transition metal layer between two chalcogen layers, and the barrier between the two TMDs is modeled by a stack of 2D layers. The interlayer distance $h$ is defined as the separation between the transition metal layers of the two TMDs and is therefore given by $h=(N_s+3)d$ with $N_s$ the number of layers in the barrier and $d=0.333$ nm the elementary distance between the different layers in the system. Therefore, when there is no barrier between the two TMDs, the interlayer distance $h$ equals 1 nm. In principle an analytic expression for $\varepsilon(q,-h/2,h/2)$ for the interaction between a charge in the transition metal layer of the top TMD and a charge in the transition metal layer of the bottom TMD can be found. This expression is very lengthy and is given in Appendix \ref{sec:appB} for a single hBN layer barrier and in the limit of no polarization in the chalcogen layers. The real space interaction potential can only be determined numerically. The interlayer interaction potential in the double trilayer model is shown in Fig. \ref{fig:interactie}(b). As expected, the additional dielectric screening effect in the chalcogen layers reduces the interaction strength. The dielectric environment above and below the heterostructure only leads to an approximately constant shift of the interaction potential.

\section{Results}
\label{sec:Results}
\begin{table}
\centering
\caption{Interlayer exciton binding energy (meV) for different TMD heterostructures modeled by two monolayers on top of a SiO$_2$ substrate ($\varepsilon_b=3.8$) with interlayer distance $h=0.6$ nm (left) and $h=1$ nm (right) (i.e. no barrier between the two TMDs). The rows and columns indicate the $n$-doped (bottom layer) and $p$-doped (top layer) materials, respectively.}
\begin{tabular}{c c c c c}
\hline
\hline
 & Mo$\text{S}_2$ & MoS$\text{e}_2$ & W$\text{S}_2$ & WS$\text{e}_2$ \\
\hline
\hline
Mo$\text{S}_2$ & 119.5/103.1 & 112.3/97.1 & 117.1/101.4 & 112.1/97.1 \\
\hline
MoS$\text{e}_2$ & 113.9/98.7 & 107.4/93.2 & 111.4/96.9 & 106.9/93.0 \\
\hline
W$\text{S}_2$ & 116.4/100.7 & 109.3/94.7 & 114.8/99.6 & 109.7/95.3 \\
\hline
WS$\text{e}_2$ & 112.7/97.7 & 106.1/92.1 & 110.9/96.5 & 106.3/92.5 \\
\hline
\hline
\end{tabular}
\label{table:bindtable}
\end{table}
\begin{table}
\centering
\caption{Average interlayer exciton in-plane interparticle distance (nm) for the same systems as in Table \ref{table:bindtable}.}
\begin{tabular}{c c c c c}
\hline
\hline
 & Mo$\text{S}_2$\hspace{10pt} & MoS$\text{e}_2$\hspace{10pt} & W$\text{S}_2$\hspace{10pt} & WS$\text{e}_2$ \\
\hline
\hline
Mo$\text{S}_2$ & 2.37/2.67\hspace{10pt} & 2.43/2.75\hspace{10pt} & 2.57/2.89\hspace{10pt} & 2.62/2.95 \\
\hline
MoS$\text{e}_2$ & 2.41/2.72\hspace{10pt} & 2.47/2.79\hspace{10pt} & 2.63/2.95\hspace{10pt} & 2.67/3.00 \\
\hline
W$\text{S}_2$ & 2.58/2.91\hspace{10pt} & 2.66/3.00\hspace{10pt} & 2.75/3.09\hspace{10pt} & 2.81/3.16 \\
\hline
WS$\text{e}_2$ & 2.61/2.93\hspace{10pt} & 2.68/3.02\hspace{10pt} & 2.79/3.13\hspace{10pt} & 2.84/3.19 \\
\hline
\hline
\end{tabular}
\label{table:distable}
\end{table}

In Table \ref{table:bindtable} (Table \ref{table:distable}) we give the binding energy (average in-plane interparticle distance) of interlayer excitons in different TMD heterostructures for two different interlayer distances: $h=0.6$ nm, which is the lower bound from ab initio predictions\cite{intdist1,intdist2}, and the above mentioned theoretical value $h=1$ nm. As such we have a range of binding energies and interparticle distances which should be relevant for experiments. For these calculations we used the parameters given in Table \ref{table:mattable}. To calculate the exciton in-plane interparticle distance we start from the electron-hole correlation function, defined as
\begin{equation}
\label{corr}
C_{eh}^{\alpha}(\vec{r}) = \braket{\Psi_{\alpha}^{exc}|\delta(\vec{r}_e-\vec{r}_h-\vec{r})|\Psi_{\alpha}^{exc}},
\end{equation}
from which we can calculate the probability to find the electron and hole at a distance $r$. For an axisymmetric system, this reduces to
\begin{equation}
\label{prob}
P_{eh}^{\alpha}(r) = 2\pi rC_{eh}^{\alpha}(r),
\end{equation}
which satisfies
\begin{equation}
\label{norm}
\int_0^{\infty}P_{eh}^{\alpha}(r)dr = 1.
\end{equation}
The average electron-hole distance is then obtained by
\begin{equation}
\label{dist}
\braket{r_{eh}^{\alpha}} = \int_0^{\infty}rP_{eh}^{\alpha}(r)dr = 2\pi\int_0^{\infty}r^2C_{eh}^{\alpha}(r)dr.
\end{equation}

It is important to note that in Tables I and II we show all the possible combinations of TMDs, including those where both TMDs are identical, for the sake of completeness. In order to optically excite interlayer excitons one needs a type-II band alignment where the energy bands of the electron TMD are shifted downwards in energy with respect to those of the hole TMD, which is predicted to occur in a wide range of TMD heterostructures\cite{type1,type2,type3,typeII,type4}. For heterostructures which do not have a type-II band alignment we artificially put the electron in one TMD and the hole in the other TMD. In theory, the necessary band alignment can always be created for any combination of TMDs using a perpendicular electric field, however the required electric field strengths may be unrealistically large depending on the band offsets. It is also possible to create interlayer excitons by external doping of the different TMDs of choice but in that case a dielectric barrier is required between the two TMDs to prevent immediate electron-hole recombination.
\begin{table}
\centering
\caption{Lattice constant ($a$) \cite{theory1}, hopping parameter ($t$) \cite{theory1}, band gap ($\Delta$) \cite{theory1}, spin splitting ($2\lambda$) \cite{spinsplitv}, and 2D polarizability ($\chi_{2\text{D}}$) \cite{berkelbach} for different TMD materials.}
\begin{tabular}{c c c c c c c}
\hline
\hline
 & $a$ (nm) & $t$ (eV) & $\Delta$ (eV) & $2\lambda$ (eV) & $\chi_{2\text{D}}$ (nm) \\
\hline
\hline
Mo$\text{S}_2$ & 0.32 & 1.10 & 1.66 & 0.15 & 8.29 \\
\hline
MoS$\text{e}_2$ & 0.33 & 0.94 & 1.47 & 0.18 & 10.34 \\
\hline
W$\text{S}_2$ & 0.32 & 1.37 & 1.79 & 0.43 & 7.58 \\
\hline
WS$\text{e}_2$ & 0.33 & 1.19 & 1.60 & 0.46 & 9.02 \\
\hline
\hline
\end{tabular}
\label{table:mattable}
\end{table}

The difference between the maximum (MoS$_2$-MoS$_2$) and minimum (WSe$_2$-MoSe$_2$) binding energy in Table \ref{table:bindtable} is 13.4 meV for $h=0.6$ nm and 11 meV for $h=1$ nm. The binding energies for $h=1$ nm are smaller than those for $h=0.6$ nm because of the reduced interaction strength. Heterostructures consisting of two TMD layers containing sulfur have noticeably larger binding energies than heterostructures consisting of two TMD layers containing selenium. The difference between the maximum (WSe$_2$-WSe$_2$) and minimum (MoS$_2$-MoS$_2$) interparticle distance in Table \ref{table:distable} is 0.47 nm for $h=0.6$ nm and 0.52 nm for $h=1$ nm. The interparticle distances for $h=1$ nm are larger than those for $h=0.6$ nm because of the reduced interaction strength. Heterostructures consisting of two TMD layers containing tungsten have noticeably larger interparticle distances than heterostructures consisting of two TMD layers containing molybdenum. Therefore we can conclude that the chalcogen atoms mostly influence the binding energy whereas the transition metal atoms mostly influence the interparticle distance. Notice also that the result for e.g. MoS$_2$-MoSe$_2$ is slightly different from that for MoSe$_2$-MoS$_2$ because of the asymmetric dielectric environment. The intralayer exciton binding energies are, respectively, 320.9 meV, 290.1 meV, 284.6 meV, and 265.1 meV for monolayer MoS$_2$, MoSe$_2$, WS$_2$, and WSe$_2$ on a SiO$_2$ substrate. These are about a factor 3 larger than the interlayer exciton binding energies. The corresponding intralayer exciton average interparticle distances are 1.02 nm, 1.05 nm, 1.21 nm, and 1.22 nm, respectively. These are about a factor 2.5 smaller than the interlayer exciton average interparticle distances.
\begin{table}
\centering
\caption{Exciton binding energy (meV) for the lowest three s-states for two different TMD heterostructures in vacuum and on top of a SiO$_2$ substrate ($\varepsilon_b=3.8$), modeled by two monolayers (left) and two trilayers with no polarization effects in the chalcogen layers (right), with interlayer distance $h=0.6$ nm and $h=1$ nm (i.e. no barrier between the two TMDs) compared to other theoretical works. Mo- and W-based TMDs are $n$-doped (bottom layer) and $p$-doped (top layer), respectively.}
\begin{tabular}{c c c c c c}
\hline
\hline
 & Substrate & State & \multicolumn{2}{c}{Current work} & Theory \\
 & & & 0.6 nm & 1 nm & \\
\hline
\hline
Mo$\text{S}_2$-WS$_2$ & Vacuum & 1s & 221 & 197 & 430 [\onlinecite{vacuum}] \\
  & & 2s & 134 & 124 & - \\
  & & 3s & 91 & 86 & - \\
  & SiO$_2$ & 1s & 117/139 & 101/120 & - \\
  & & 2s & 56/65 & 51/59 & - \\
  & & 3s & 32/36 & 30/34 & - \\
\hline
MoS$\text{e}_2$-WSe$_2$ & Vacuum & 1s & 195 & 175 & 320 [\onlinecite{vacuum}] \\
  & & 2s & 123 & 114 & - \\
  & & 3s & 86 & 81 & - \\
  & SiO$_2$ & 1s & 107/124 & 93/108 & 173 [\onlinecite{sio2}] \\
  & & 2s & 54/62 & 49/57 & 69 [\onlinecite{sio2}] \\
  & & 3s & 32/36 & 30/33 & 35 [\onlinecite{sio2}] \\
\hline
\hline
\end{tabular}
\label{table:comptable}
\end{table}

In Table \ref{table:comptable} we compare the interlayer exciton binding energies for two different TMD heterostructures with the other theoretical works mentioned in the Introduction. We find considerably smaller binding energies which is possibly due to the fact that other models for the interaction potential are used in the other theoretical works. The agreement is better for higher excited states, which is because the binding energy converges to 0 in this limit, and for smaller interlayer distances. In Ref. [\onlinecite{vacuum}] interlayer distances between 0.6 nm and 0.65 nm were used whereas in Ref. [\onlinecite{sio2}] an interlayer distance of 0.645 nm was used. Little to no details on the interaction potential are given in Ref. [\onlinecite{vacuum}] but the authors claim that their large interlayer exciton binding energies, which are only 20\% smaller than the intralayer exciton binding energies, are the result of reduced out-of-plane screening. However, we found in Fig. \ref{fig:interactie} that the interlayer interaction potential is considerably weaker than what would be expected from a simple substitution $r\rightarrow\sqrt{r^2+h^2}$ in the intralayer interaction potential because there is a screening effect in both TMDs. In Ref. [\onlinecite{sio2}] a model similar to ours is used for the interlayer interactions, except that the TMDs are modeled by homogeneous slabs with a certain thickness and constant dielectric constant, meaning that there can be a spacing between the charge carriers and the substrate, whereas we model the TMDs by strictly 2D materials with a 2D polarizability. To facilitate comparison we have therefore also done the calculations for the case in which the TMDs are modeled by trilayers without polarization effects in the chalcogen layers, meaning that there is a spacing of 0.666 nm between the transition metal layer and the SiO$_2$ substrate, in accordance with Fig. \ref{fig:schema}(b). These results are also shown in Table \ref{table:comptable} and we find that the interlayer exciton binding energies are larger due to the reduced influence of the substrate and as such the results are closer to those of Ref. [\onlinecite{sio2}].
\begin{figure}
\centering
\includegraphics[width=8.5cm]{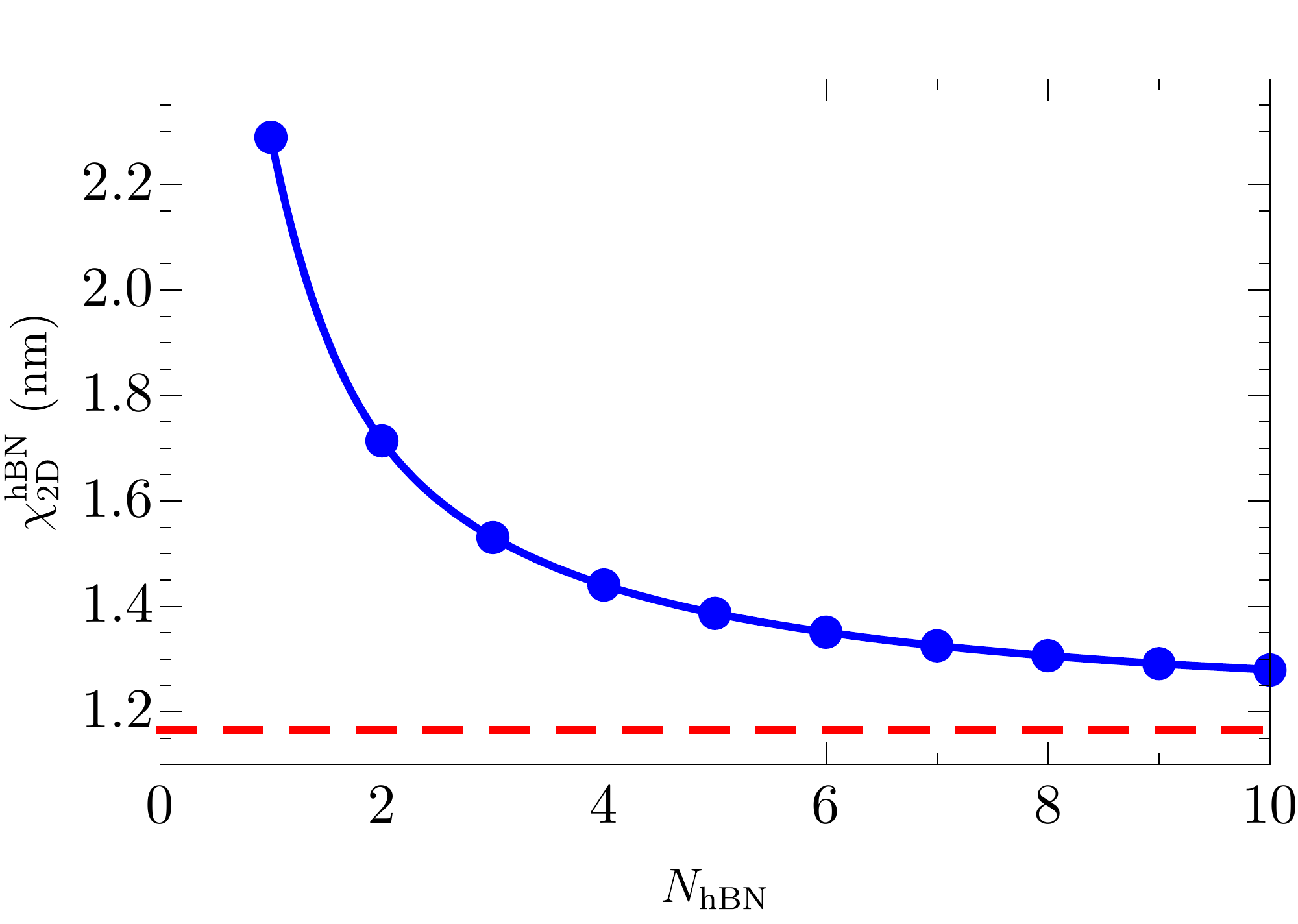}
\caption{(Color online) 2D polarizability of the hBN layers for which the interaction potential is identical to that when a homogeneous barrier is present between the TMDs, which we model by monolayers, with $\varepsilon_r=4.5$ and $\varepsilon_z=1$, as a function of the number of hBN layers. We take $\varepsilon_b=\varepsilon_t=1$ and $\chi_{2\text{D}}^b=\chi_{2\text{D}}^t=8$ nm. The thickness of the homogeneous barrier is modified according to the number of hBN layers. The red dashed line indicates the in-plane 2D polarizability $\chi_{2\text{D}}^{\text{hBN}}=d(\varepsilon_r-1)$. The blue curve is a guide to the eye.}
\label{fig:chi}
\end{figure}

There are two possible ways of modeling the presence of hBN as a barrier: as a homogeneous 3D slab with a relative dielectric constant $\varepsilon_s$ or as a stack of 2D layers with 2D polarizability $\chi_{2\text{D}}^{\text{hBN}}$. The relation between these two parameters is approximately given by $\chi_{2\text{D}}^{\text{hBN}}=d(\varepsilon_s-1)$\cite{screening3}. When we calculate the interlayer interaction potential in both models we find that they can never be identical when hBN is assumed to be isotropic in the 3D model. It turns out that we need to put the out-of-plane relative dielectric constant equal to 1 in order to have equal interaction potentials in the two models. In Fig. \ref{fig:chi} we show the 2D polarizability for which the interaction potential in the 2D model is identical to that in the 3D model with $\varepsilon_z=1$ and $\varepsilon_r=4.5$ as a function of the number of hBN layers. This shows that for an increasing number of hBN layers the equivalent 2D polarizability converges to the result found using the formula of Ref. [\onlinecite{screening3}]. For a finite number of layers the 2D polarizability is always larger than this limiting value, with a maximum difference of a factor 2 for a single hBN layer.
\begin{table}
\centering
\caption{Exciton binding energy (meV) for two different TMD heterostructures in two different dielectric environments for 1 up to 3 layers of hBN between the two TMDs which are modeled by trilayers with (right) and without (left) polarization effects in the chalcogen layers with $\chi_{2\text{D}}^X=\chi_{2\text{D}}^M/4$. We use $\varepsilon_b=3.8$ for SiO$_2$ and ($\varepsilon_r=4.5,\varepsilon_z=1$) for the hBN substrates. For the hBN layers in the barrier we use $\chi_{2\text{D}}^{\text{hBN}}=1.17$ nm. Mo- and W-based TMDs are $n$-doped (bottom layer) and $p$-doped (top layer), respectively.}
\begin{tabular}{c c c c c c}
\hline
\hline
 & Substrate $t/b$ & $N_{\text{hBN}}=1$ & $N_{\text{hBN}}=2$ & $N_{\text{hBN}}=3$ \\
\hline
\hline
Mo$\text{S}_2$-WS$_2$ & Vacuum/SiO$_2$ & 103/83 & 90/74 & 81/68 \\
  & hBN/hBN & 133/107 & 118/97 & 107/88 \\
\hline
MoS$\text{e}_2$-WSe$_2$ & Vacuum/SiO$_2$ & 94/75 & 83/68 & 75/62 \\
  & hBN/hBN & 121/96 & 108/87 & 98/81 \\
\hline
\hline
\end{tabular}
\label{table:subtable}
\end{table}

In Table \ref{table:subtable} we show the interlayer exciton binding energy for two different heterostructures, modeled by two trilayers with and without polarization effects in the chalcogen layers, in two different dielectric environments and for a different number of hBN layers between the two TMDs. As expected, the binding energy decreases with increasing number of layers in the barrier and when there are polarization effects in the chalcogen layers, with the latter effect being stronger than the former. Even though the in-plane dielectric constant of hBN is larger than that of SiO$_2$, the interlayer exciton binding energy of a system with hBN both above and below the heterostructure is larger than that of a system with vacuum (SiO$_2$) above (below) the heterostructure. This is because the out-of-plane dielectric constant of hBN is smaller than that of SiO$_2$.

\begin{figure}
\centering
\includegraphics[width=8.5cm]{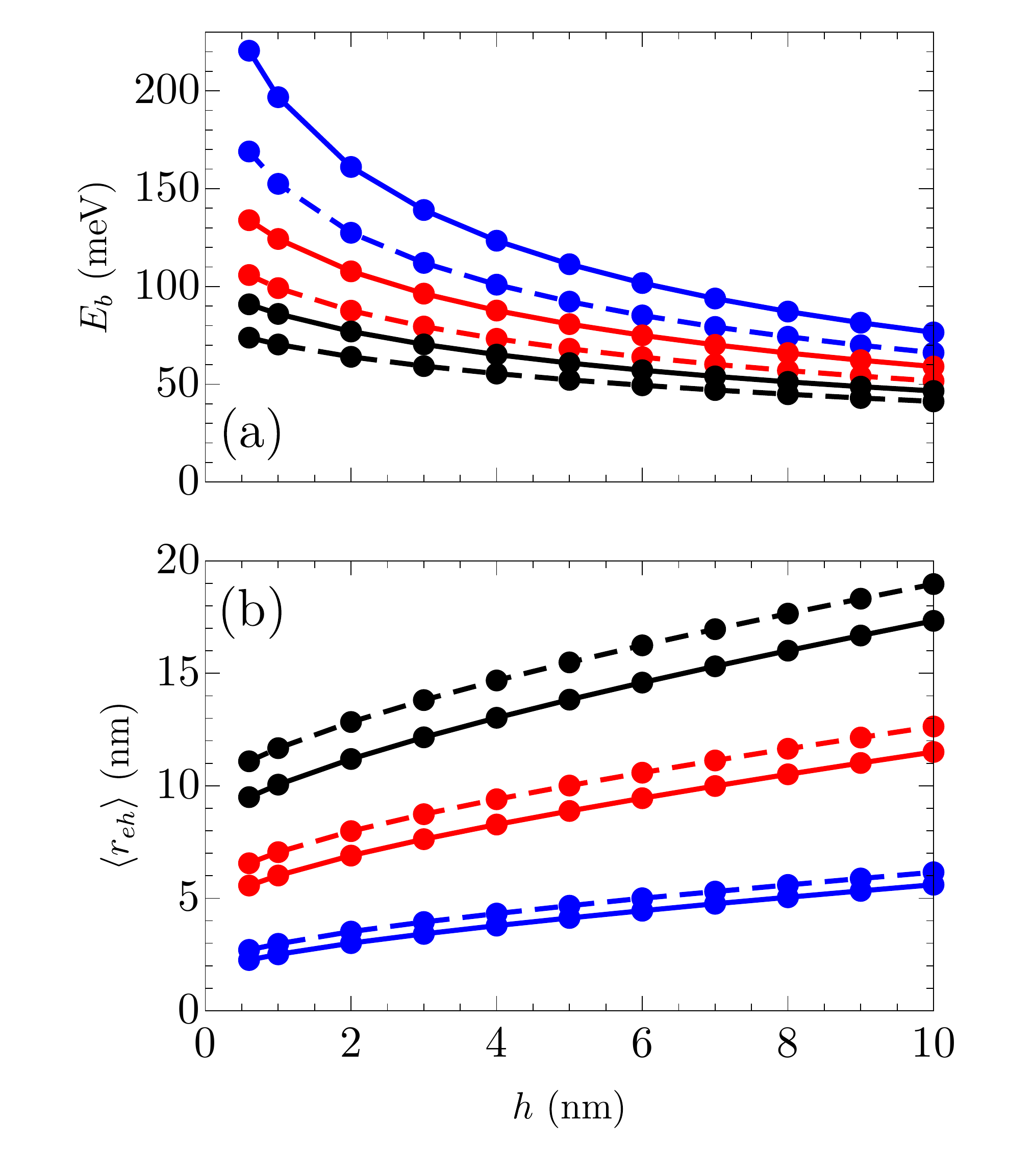}
\caption{(Color online) Binding energy (a) and average in-plane interparticle distance (b) for  the 1s (blue), 2s (red), and 3s (black) interlayer excitons in a MoS$_2$-WS$_2$ heterostructure in vacuum, modeled by two monolayers (solid) and two trilayers with $\chi_{2\text{D}}^X=\chi_{2\text{D}}^M/4$ (dashed). We take no barrier between the two TMDs. Mo- and W-based TMDs are $n$-doped (bottom layer) and $p$-doped (top layer), respectively.}
\label{fig:hplot}
\end{figure}

In Fig. \ref{fig:hplot} we show the interlayer exciton binding energy (a) and average in-plane interparticle distance (b) as a function of the interlayer distance and we compare the monolayer and the trilayer model for the TMDs. The binding energy decreases in both models with increasing interlayer distance due to the reduced interaction strength, with the binding energy at $h=10$ nm being more than twice as small as the value at $h=1$ nm for the ground state. The additional polarization in the chalcogen layers in the double trilayer model reduces the binding energy by an amount in the order of tens of meV, with the effect being more pronounced at small interlayer distances. For higher excited states, which have smaller binding energy, the effect of the chalcogen layers is less pronounced. Correspondingly, the average interparticle distance of the interlayer exciton increases with increasing interlayer distance, reaching more than twice the value of $h=1$ nm at $h=10$ nm for the ground state. The polarization in the chalcogen layers increases the average interparticle distance by about 0.5 nm, 1nm, and 1.5 nm for the 1s, 2s, and 3s exciton, respectively. This means that the chalcogen layers lead to a larger absolute increase in the average interparticle distance of higher excited states, although the relative increase in interparticle distance is smaller for higher excited states. Remarkably, in absolute terms, the effect of this additional polarization on the interparticle distance is approximately independent on the interlayer distance.

\begin{figure}
\centering
\includegraphics[width=8.5cm]{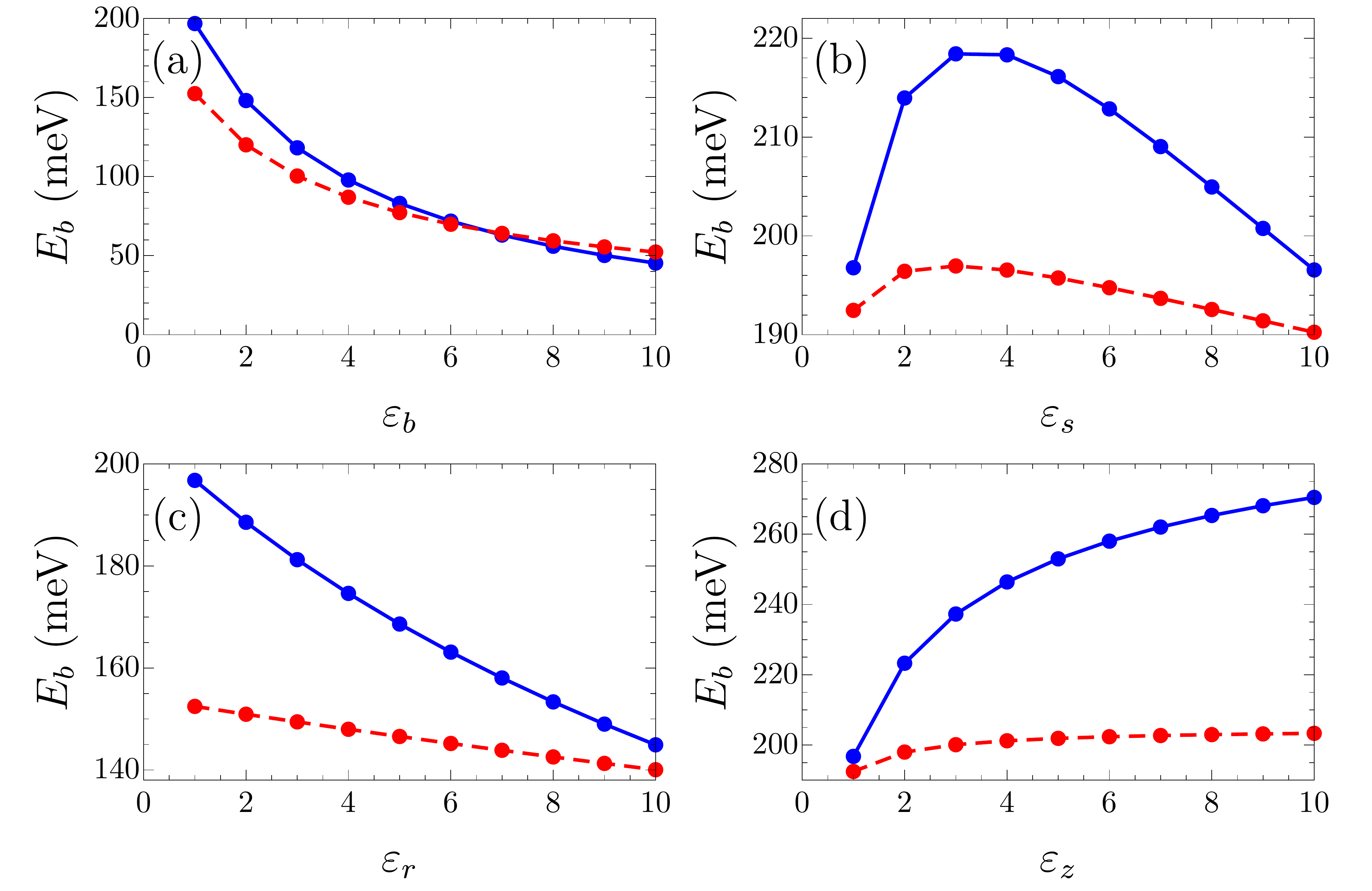}
\caption{(Color online) Binding energy for interlayer excitons in a MoS$_2$-WS$_2$ heterostructure with interlayer distance $h=1$ nm, modeled by two monolayers (blue, solid) and two trilayers with $\chi_{2\text{D}}^X=\chi_{2\text{D}}^M/4$ (red, dashed) as a function of $\varepsilon_b$ (a), $\varepsilon_s$ (b), $\varepsilon_r$ (c), and $\varepsilon_z$ (d). The dielectric constants which are not varied are set to 1 for each figure. The red dashed curves in (b) and (d) are shifted upwards by 40 meV for clarity. Mo- and W-based TMDs are $n$-doped (bottom layer) and $p$-doped (top layer), respectively.}
\label{fig:epsplot}
\end{figure}

We show the dependencies of the interlayer exciton binding energy on the different relative dielectric constants of the system in Fig. \ref{fig:epsplot}. The exciton binding energy decreases by a factor 4 when the dielectric constant of the substrate below the TMD heterostructure increases from 1 to 10, as shown in Fig. \ref{fig:epsplot}(a). When the polarization in the chalcogen layers is taken into account in the double trilayer model the decrease in binding energy is limited to a factor 3 for the same dielectric constant range. At low dielectric constants this additional polarization leads to a decrease in binding energy. However, for dielectric constants above $\varepsilon_b\approx7$ the additional polarization leads to an increase in binding energy. This can be explained due to the fact that, although the chalcogen layers themselves weaken the interactions, they also screen the effect of the substrate. When the effect of the substrate is stronger than that of the chalcogen layers, i.e. for large values of $\varepsilon_b$, this screening of the substrate can enhance the total interaction strength.

We find a remarkable dependence on the isotropic dielectric constant $\varepsilon_s=\varepsilon_r=\varepsilon_z$ of the barrier between the two TMD layers, as shown in Fig. \ref{fig:epsplot}(b). At first the binding energy increases with increasing $\varepsilon_s$, before reaching a maximum at $\varepsilon_s\approx4$ after which it starts to decrease. When the additional polarization in the chalcogen layers is taken into account this dependence changes quantitatively, the binding energy is reduced, the dependence on $\varepsilon_s$ is less pronounced, and the maximum binding energy is reached at a slightly smaller value of $\varepsilon_s$, but qualitatively it remains the same. To gain more insight into this behavior we also study the anisotropic case. The $\varepsilon_r$-dependence, as shown in Fig. \ref{fig:epsplot}(c), is similar to the $\varepsilon_b$-dependence but it is less strong. At some value of $\varepsilon_r$ between 12 and 13 the additional polarization in the chalcogen layers again leads to an increase in the exciton binding energy. It is not entirely clear what physical mechanism is behind this increase. Finally, we show the $\varepsilon_z$-dependence of the exciton binding energy in Fig. \ref{fig:epsplot}(d). In contrast to what might be expected, we find that the binding energy increases as a function of $\varepsilon_z$. In the limit of large $\varepsilon_z$ it converges to a fixed value. When the polarization in the chalcogen layers is taken into account this convergence occurs at smaller values of $\varepsilon_z$. We find limiting values of $E_b=315$ meV and $E_b=165$ meV in the absence and presence of the chalcogen layers, respectively. We can conclude that $\varepsilon_r$ and $\varepsilon_z$ are in competition with each other for the $\varepsilon_s$-dependence of the binding energy. It is the increase as a function of $\varepsilon_z$ which causes the binding energy in (b) to increase for small values of $\varepsilon_s$, whereas the decrease as a function of $\varepsilon_r$ and the convergence at large $\varepsilon_z$ lead to the subsequent decrease in binding energy.
\begin{figure}
\centering
\includegraphics[width=8.5cm]{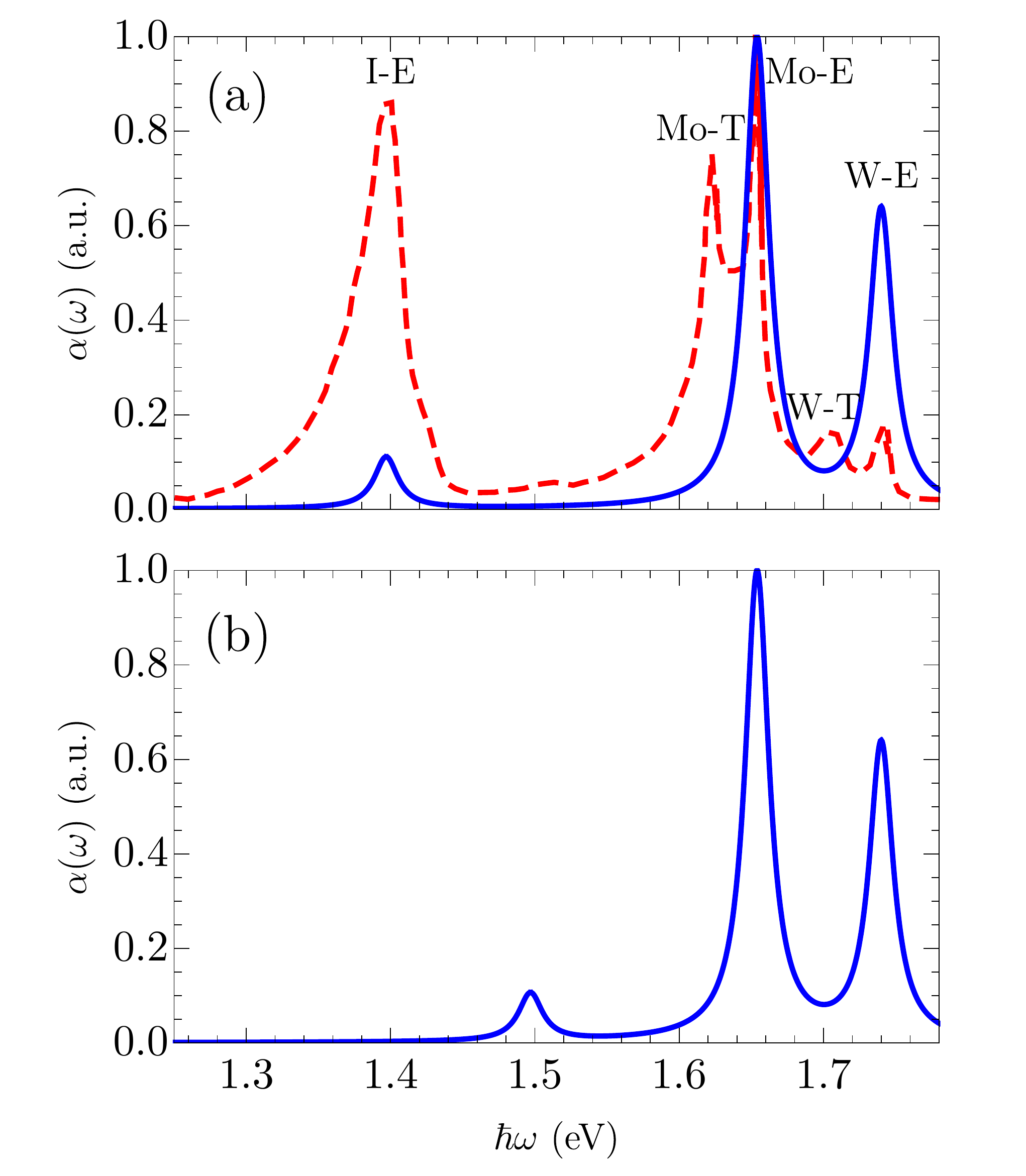}
\caption{(Color online) Excitonic absorbance spectra for a MoSe$_2$-WSe$_2$ heterostructure modeled by two monolayers on a SiO$_2$ substrate ($\varepsilon_b=3.8$) with vacuum on top, in the absence (a) and presence (b) of a perpendicular electric field of $-$0.1 V/nm and with interlayer distance $h=1$ nm (i.e. no barrier between the two TMDs). We used a broadening of $\gamma=10$ meV. The red dashed curve is the experimental photoluminescence result from Ref. [\onlinecite{exp3}]. The interlayer (I) and intralayer (Mo/W) exciton (E) and trion (T) peaks are indicated on the figure. Mo- and W-based TMDs are $n$-doped (bottom layer) and $p$-doped (top layer), respectively.}
\label{fig:plplot}
\end{figure}
\begin{figure}
\centering
\includegraphics[width=8.5cm]{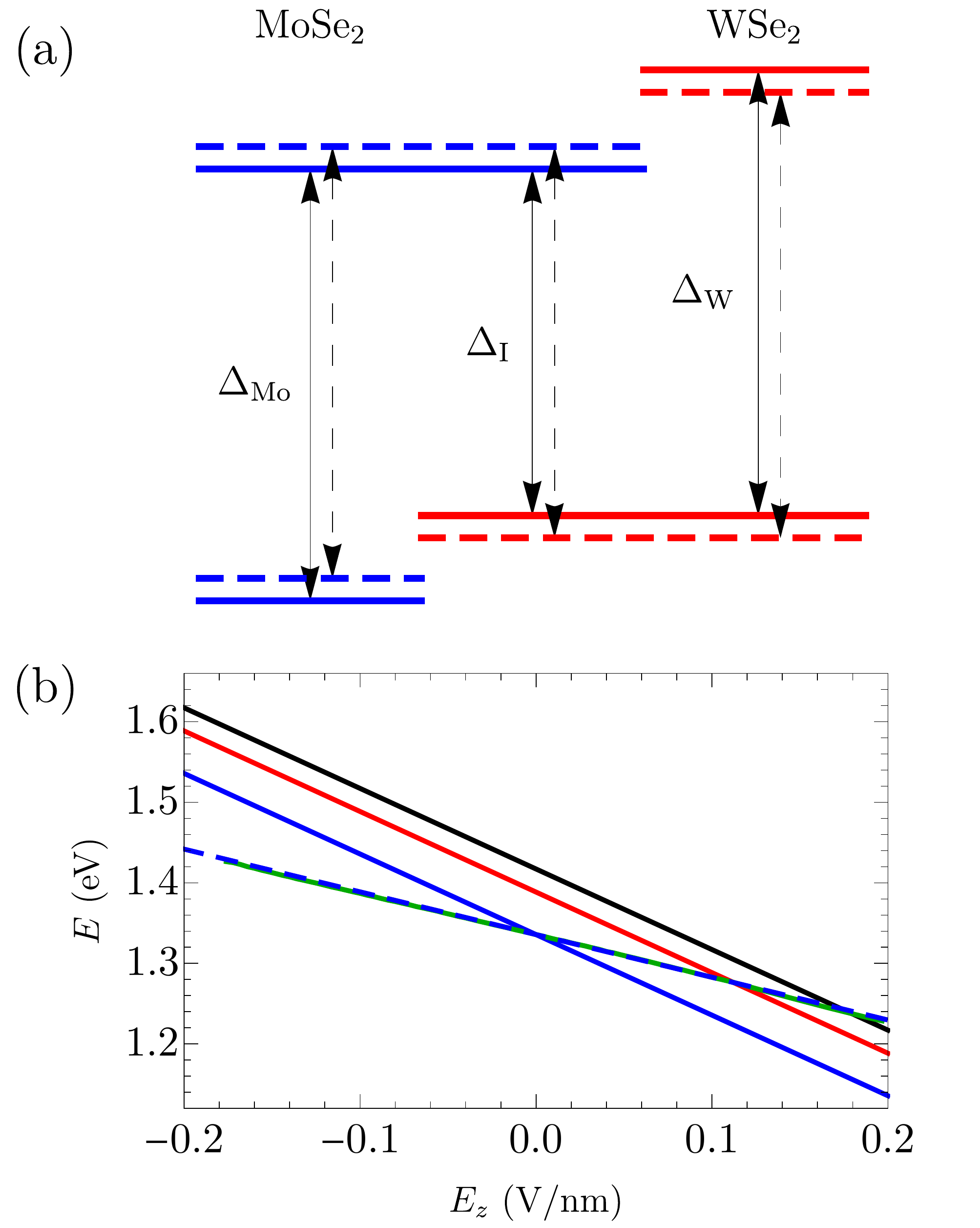}
\caption{(Color online) (a) Schematic of the band diagram of a MoSe$_2$-WSe$_2$ heterostructure with (dashed) and without (solid) a perpendicular electric field. $\Delta_{\text{Mo}}$ ($\Delta_{\text{W}}$) indicates the intralayer exciton band gap in MoSe$_2$ (WSe$_2$) and $\Delta_{\text{I}}$ indicates the interlayer exciton band gap. (b) 1s (blue), 2s (red), and 3s (black) interlayer exciton energy as a function of the perpendicular electric field pointing from MoSe$_2$ to WSe$_2$ in the case of a hBN substrate above and below the heterostructure. We model the TMDs by monolayers. The green curve is the experimental result for the 1s interlayer exciton from Ref. [\onlinecite{Eexp}]. The dashed blue curve is our theoretical result when we reduce the interlayer distance from 1 nm to 0.53 nm.}
\label{fig:Ediaplot}
\end{figure}

When both the excitonic energy spectrum as well as the wave functions are known we can also calculate the absorbance spectrum using the formula\cite{absorbform}
\begin{equation}
\label{absorbance}
\alpha(\omega) \propto \frac{1}{\omega}\text{Im}\left(\sum_j\frac{|\mathcal{P}_0|^2|\phi_{c,v}^{e,h,j}(0)|^2}{E_j-\hbar\omega-i\gamma}\right),
\end{equation}
with $E_j$ the exciton energy of state $j$, $\phi_{c,v}^{e,h,j}$ the corresponding dominant component of the exciton wave function, $\hbar\omega$ the photon energy, $\gamma$ the broadening of the peaks and where $\mathcal{P}_0=2m_0at/\hbar$ is the coupling strength with optical fields of circular polarization evaluated at the band edges\cite{theory1}. The result is shown in Fig. \ref{fig:plplot}. The highest energy peak corresponds to intralayer excitons in the WSe$_2$ layer. The peak next to it corresponds to intralayer excitons in the MoSe$_2$ layer. The small low-energy peak corresponds to interlayer excitons. We have modified our band gap parameters in order to align the different peaks with those from the experimental results. As such we find an offset between the conduction bands of the two TMDs of 515 meV and a valence band offset of 453 meV. These results are indeed larger than the lower bounds of 310 meV and 230 meV for the conduction band offset and valence band offset, respectively, which were found in Ref. [\onlinecite{exp3}]. The additional peaks which are present in the experimental results correspond to trions, which are not considered in our calculations. There are considerable differences in the intensities of the interlayer exciton and the WSe$_2$ intralayer exciton peaks between our absorbance spectrum and the experimental photoluminescence spectrum. The difference lies in the degree of occupation of the different excitonic states. This depends on multiple factors such as the temperature, laser power, recombination times, \ldots, and is therefore difficult to predict.

Finally, we also show the results in the presence of a perpendicular electric field of $-$0.1 V/nm, which is added as a constant term to the diagonal elements of the exciton Hamiltonian \eqref{hamtot}. The intralayer exciton peaks are unaffected, but the interlayer exciton peak shifts upwards in energy by about 0.1 eV. Mathematically, this can be understood since the electric field shifts the energy bands of the two TMDs with respect to each other, thus reducing the interlayer exciton band gap, while the intralayer exciton band gaps remain the same. This is shown schematically in Fig. \ref{fig:Ediaplot}(a). Physically, this is because the interlayer excitons form an electric dipole pointing (partly) in the perpendicular direction and as such couple to a perpendicular electric field. In this case the electric field is oriented opposite to the interlayer exciton dipole moment and as such the interlayer exciton energy is increased. This is very different for intralayer excitons which have an electric dipole pointing completely in the material plane and as such do not couple to a perpendicular electric field. This effect was also found in Ref. [\onlinecite{Eexp}] in which a bottom and top gate were placed on the hBN substrate above and below the material, respectively. In Fig. \ref{fig:Ediaplot}(b) we compare these experimental results with our results. The interlayer exciton energy depends linearly on the perpendicular electric field, which corresponds with the energy of a electric dipole in an electric field. However, we find that the slopes of our curves, which are determined by the interlayer distance, do not agree with the slope found in the experimental results. When we take the interlayer distance as a fitting parameter we find excellent agreement with the experimental results when we take an interlayer distance of 0.53 nm, as opposed to our assumed interlayer distance of 1 nm. This indicates that interlayer exciton formation may cause the electrons and holes to be pulled out of the transition metal layer. This effect is not taken into account in our single-particle Hamiltonian \eqref{singelhame} which describes a strictly 2D system. A similar experiment was carried out in Ref. [\onlinecite{exp3}], in which vacuum (SiO$_2$) was placed above (below) the heterostructure and where only a top gate and a backgate were used, which is more difficult to model theoretically. They found non-linear behavior as a function of the backgate potential, however they mention explicitly that the use of top and bottom gates may elucidate this phenomenon.

\section{Summary and conclusion}
\label{sec:Summary and conclusion}

In this paper, we studied interlayer excitons in TMD heterostructures. We started from the single-particle Dirac Hamiltonian to construct a four-band exciton Hamiltonian and we solved the corresponding eigenvalue equation using the finite element method. Starting from Gauss's law in dielectrics, we derived an expression for the electron-hole interaction potential taking into account the effects of the different dielectric environments and the polarization effects in the transition metal layer and in the chalcogen layers of the TMD layers. We have modeled the barrier between the two TMD layers both by a 3D slab and by a stack of 2D layers and found that the two models can only be mapped onto each other when the dielectric constant perpendicular to the layers of the barrier is taken to be $\varepsilon_z=1$.

We investigated the effect of additional polarization in the chalcogen layers and found that this effect is most pronounced when the interlayer exciton binding energy is large, i.e. at small interlayer distances and/or small substrate or barrier dielectric constants, meaning that the average in-plane interparticle distance is small. In general this extra polarization effect reduces the exciton binding energy. However, when there are very strong substrate effects present in the system it can lead to an increase in binding energy because it screens the substrates.

Furthermore, we investigated the dependence of the exciton binding energy on the different dielectric constants of the dielectric environment and found remarkable behavior, i.e. an initial increase followed by a steady decrease, as a function of the dielectric constant of the barrier between the two layers. We could link this behavior to the presence of two competing effects: a decrease of the binding energy as a function of the in-plane dielectric constant and an increase as a function of the out-of-plane dielectric constant, although why this latter effect occurs remains an open question.

Finally, we calculated the excitonic absorbance spectrum and compared it with recent experimental results. By doing so we were able to obtain the band offsets for both the conduction and the valence band. We also investigated the effect of a perpendicular electric field on the absorbance spectrum and found that it shifts the interlayer exciton peak linearly, which was in perfect agreement with experiment if we changed our interlayer distance from 1 nm to 0.53 nm, while the intralayer exciton peaks remain unaffected.

\section{Acknowledgments}

This work was supported by the Research Foundation of Flanders (FWO-Vl) through an aspirant research grant for MVDD and by the FLAG-ERA project TRANS-2D-TMD.

\appendix

\section{Derivation of the interlayer interaction potential}
\label{sec:appA}

In order to find an expression for the interlayer interaction potential we start from Gauss' law: $\vec{\nabla}.\vec{D}=n_{ext}$, with $n_{ext}$ the charge density of an external point charge located at $(\vec{r}',z')$ with charge $q_1$. The displacement field $\vec{D}$ is given by $\vec{D}=\varepsilon_0\vec{E}+\vec{P}$, with $\vec{E}$ the electric field and $\vec{P}=\chi\varepsilon_0\vec{E}$ the polarization density, with $\chi$ the polarizability. For homogeneous 3D dielectrics this simplifies to $\vec{D}=\tilde{\varepsilon}\vec{E}$, with $\tilde{\varepsilon}$ the dielectric tensor of the material. Using $\vec{E}=-\vec{\nabla}\phi(\vec{r}-\vec{r}',z,z')$, with $\phi(\vec{r}-\vec{r}',z,z')$ the electrostatic potential, Gauss's law becomes
\begin{equation}
\label{gausslaw}
\begin{split}
&\left(\varepsilon_r^i\left(\frac{\partial^2}{\partial x^2}+\frac{\partial^2}{\partial y^2}\right)+\frac{\partial}{\partial z}\left(\varepsilon_z^i\frac{\partial}{\partial z}\right)\right)\phi(\vec{r}-\vec{r}',z,z') \\
&\hspace{115pt}= -\frac{q_1}{\varepsilon_0}\delta\left(\vec{r}-\vec{r}'\right)\delta\left(z-z'\right),
\end{split}
\end{equation}
with $\varepsilon_r^i$ and $\varepsilon_z^i$ the in-plane and out-of-plane relative dielectric constants of the homogeneous 3D region $i$.
For 2D materials such as the layers of the TMDs as well as the layers of a layered substrate or barrier such as hexagonal boron nitride (hBN) there will only be an induced charge density in the material plane\cite{trilaag}, i.e. $\chi(z)=\chi_{2\text{D}}\delta(z-z_0)$ for a 2D material located at $z_0$, where $\chi_{2\text{D}}$ has the dimensions of length as opposed to the dimensionless 3D polarizability $\chi$.  As a consequence it is no longer possible to write $\vec{D}=\tilde{\varepsilon}\vec{E}$ and it is this nonlocal dielectric screening which leads to the interaction potential \eqref{interpot} inside a single layer. Gauss's law becomes $\varepsilon_0\vec{\nabla}.\vec{E}=n_{ext}+\sum_{j=1}^Nn_{ind}^j$, with $N$ the number of 2D layers in the system and with
\begin{equation}
\label{inducedcharge}
\begin{split}
n_{ind}^j &= -\vec{\nabla}.\vec{P}^j = \varepsilon_0\chi_{2\text{D}}^j\vec{\nabla}.\left(\delta\left(z-z_j\right)\vec{\nabla}\phi(\vec{r}-\vec{r}',z,z')\right) \\
&= \varepsilon_0\chi_{2\text{D}}^j\bigg(\delta\left(z-z_j\right)\left(\frac{\partial^2}{\partial x^2}+\frac{\partial^2}{\partial y^2}\right)+\delta\left(z-z_j\right)\frac{\partial^2}{\partial z^2} \\
&\hspace{43pt}+\left(\frac{\partial}{\partial z}\delta\left(z-z_j\right)\right)\frac{\partial}{\partial z}\bigg)\phi(\vec{r}-\vec{r}',z,z')
\end{split}
\end{equation}
the induced charge density in the layer located at $z_j$ with 2D polarizability $\chi_{2\text{D}}^j$. Adding the above induced charge densities to Eq. \eqref{gausslaw} and performing an in-plane 2D Fourier transform over $\vec{r}-\vec{r}'$ gives the equation
\begin{equation}
\label{gausslawfourier}
\begin{split}
&\frac{\partial}{\partial z}\left(\varepsilon_z^i\frac{\partial}{\partial z}\phi_{\vec{q}}(z,z')\right)-\varepsilon_r^iq^2\phi_{\vec{q}}(z,z') = -\frac{q_1}{A\varepsilon_0}\delta\left(z-z'\right) \\
&+\sum_{j=1}^N\chi_{2\text{D}}^j\bigg(q^2\delta\left(z-z_j\right)-\delta\left(z-z_j\right)\frac{\partial^2}{\partial z^2}\\
&\hspace{50pt}-\left(\frac{\partial}{\partial z}\delta\left(z-z_j\right)\right)\frac{\partial}{\partial z}\bigg)\phi_{\vec{q}}(z,z'),
\end{split}
\end{equation}
with $A$ the area of the system. This equation has to be solved for each homogeneous 3D region $i$ in the system\cite{transfer}. In these regions the right hand side of the above equation vanishes and the solutions are given by
\begin{equation}
\label{regionsol}
\phi_{\vec{q}}^i(z,z') = A_i(z')e^{\sqrt{\varepsilon_r^i/\varepsilon_z^i}qz}+B_i(z')e^{-\sqrt{\varepsilon_r^i/\varepsilon_z^i}qz},
\end{equation}
with $A_i$ and $B_i$ integration constants. The external and induced charge densities are located in the layers between the different homogeneous 3D regions and as such will only enter in the boundary conditions relating the different piecewise solutions $\phi_{\vec{q}}^i(z)$ at the interfaces at $z_i$ ($z'$ is equal to one of the $z_i$ because we assume that the external charge is located in one of the TMDs). We take region $i\ (i=1,\ldots,N+1)$ to be located between $z_{i-1}$ and $z_i$, implying that $z_0=-\infty$ and $z_{N+1}=+\infty$. Furthermore, we have to impose $B_1=A_{N+1}=0$ to avoid divergences. The boundary conditions are given by
\begin{equation}
\label{boundary}
\begin{split}
\phi_{\vec{q}}^{i+1}(z_i,z') &= \phi_{\vec{q}}^i(z_i,z') \\
\varepsilon_z^{i+1}\frac{\partial}{\partial z}\phi_{\vec{q}}^{i+1}(z_i,z') &= \varepsilon_z^i\frac{\partial}{\partial z}\phi_{\vec{q}}^i(z_i,z')+q^2\chi_{2\text{D}}^i\phi_{\vec{q}}^i(z_i,z') \\
&\hspace{10pt}-\frac{q_1}{A\varepsilon_0}\delta_{z',z_i}.
\end{split}
\end{equation}
Notice that the last two terms on the right hand side of Eq. \eqref{gausslawfourier} cancel each other. The interaction potential between the external charge $q_1$ and a charge $q_2$ in region $i$ can in general be written as
\begin{equation}
\label{interactiefourier}
V(q,z,z') = q_2\phi_{\vec{q}}^i(z,z') = \frac{q_1q_2}{2Aq\varepsilon_0\varepsilon(q,z,z')}
\end{equation}
with $\varepsilon(q,z,z')$ a relative dielectric function. The real space interaction potential can then be found by performing the inverse 2D Fourier transform, which gives
\begin{equation}
\label{interactiereeel}
\begin{split}
V(|\vec{r}-\vec{r}'|,z,z') &= \frac{A}{4\pi^2}\int d^2qV(q,z,z')e^{i\vec{q}.(\vec{r}-\vec{r}')} \\
&= \frac{q_1q_2}{4\pi\varepsilon_0}\int_0^{\infty}dq\frac{J_0(q|\vec{r}-\vec{r}'|)}{\varepsilon(q,z,z')},
\end{split}
\end{equation}
with $J_0$ the zeroth order Bessel function of the first kind. The relative dielectric function for a TMD heterostructure is discussed in more detail in the main text.

\begin{widetext}
\section{Dielectric function in the presence of a layered hBN barrier}
\label{sec:appB}

The dielectric function of the system shown in Fig. \ref{fig:schema}(b) can be found by solving the Poisson equation. For the case of a single hBN layer barrier and in the limit of no polarization of the chalcogen layers we find the following expression:
\begin{equation}
\label{diefunhbn}
\begin{split}
&\varepsilon\left(q,-\frac{h}{2},\frac{h}{2}\right) = \\
&\bigg(-e^{-4dq}(1-\varepsilon_b)(1-\varepsilon_t)(2-\chi_{2\text{D}}^bq)(2-\chi_{2\text{D}}^tq)(2-\chi_{2\text{D}}^{\text{hBN}}q)+e^{12dq}(1+\varepsilon_b)(1+\varepsilon_t)(2+\chi_{2\text{D}}^bq)(2+\chi_{2\text{D}}^tq)(2+\chi_{2\text{D}}^{\text{hBN}}q) \\
&+4\chi_{2\text{D}}^bq(1-\varepsilon_t)(1+\varepsilon_b-\varepsilon_b\chi_{2\text{D}}^{\text{hBN}}q)+4\chi_{2\text{D}}^tq(1-\varepsilon_b)(1+\varepsilon_t-\varepsilon_t\chi_{2\text{D}}^{\text{hBN}}q)-2\chi_{2\text{D}}^b\chi_{2\text{D}}^tq^2(2-2\varepsilon_b\varepsilon_t-(\varepsilon_b+\varepsilon_t-2\varepsilon_b\varepsilon_t)\chi_{2\text{D}}^{\text{hBN}}q) \\
&+e^{8dq}\Big(4\chi_{2\text{D}}^bq(1+\varepsilon_t)(1-\varepsilon_b-\varepsilon_b\chi_{2\text{D}}^{\text{hBN}}q)+4\chi_{2\text{D}}^tq(1+\varepsilon_b)(1-\varepsilon_t-\varepsilon_t\chi_{2\text{D}}^{\text{hBN}}q)-2\chi_{2\text{D}}^b\chi_{2\text{D}}^tq^2(2-2\varepsilon_b\varepsilon_t \\
&-(\varepsilon_b+\varepsilon_t+2\varepsilon_b\varepsilon_t)\chi_{2\text{D}}^{\text{hBN}}q)\Big)+e^{4dq}\Big(4\chi_{2\text{D}}^{\text{hBN}}q(2-2\varepsilon_b\varepsilon_t+(\varepsilon_b-\varepsilon_t)\chi_{2\text{D}}^bq)-4(\varepsilon_b+\varepsilon_t)\chi_{2\text{D}}^b\chi_{2\text{D}}^tq^2 \\
&+2q^2\chi_{2\text{D}}^t\chi_{2\text{D}}^{\text{hBN}}(2(\varepsilon_t-\varepsilon_b)+(3\varepsilon_b\varepsilon_t-1)\chi_{2\text{D}}^bq)\Big)\bigg)/\bigg(8(1-\varepsilon_b+e^{4dq}(1+\varepsilon_b))(1-\varepsilon_t+e^{4dq}(1+\varepsilon_t))\bigg).
\end{split}
\end{equation}
\end{widetext}


\begin{thebibliography}{10}

\bibitem{mak1}
K. F. Mak, C. Lee, J. Hone, J. Shan, and T. F. Heinz,  Phys. Rev. Lett. {\bf 105}, 136805 (2010).

\bibitem{splendiani}
A. Splendiani, L. Sun, Y. Zhang, T. Li, J. Kim, Ch. Chim, G. Galli, and F. Wang, Nano Lett. {\bf 10}, 1271 (2010).

\bibitem{mak2}
K. F. Mak, K. He, J. Shan, and T. F. Heinz, Nat. Nanotechnol. {\bf 7}, 494 (2012).

\bibitem{zeng}
H. Zeng, J. Dai, W. Yao, D. Xiao, and X. Cui, Nat. Nanotechnol. {\bf 7}, 490 (2012).

\bibitem{cao}
T. Cao, G. Wang, W. Han, H. Ye,	Ch. Zhu, J. Shi, Q. Niu, P. Tan, E. Wang, B. Liu, and J. Feng, Nat. Commun. {\bf 3}, 887 (2012).

\bibitem{ross}
J. S. Ross, S. Wu, H. Yu,	N. J. Ghimire,	A. M. Jones, G. Aivazian, J. Yan, D. G. Mandrus, Di Xiao, W. Yao, and X. Xu, Nature Commun. {\bf 4}, 1474 (2013).

\bibitem{korn}
T. Korn, S. Heydrich, M. Hirmer, J. Schmutzler, and C. Schller, Appl. Phys. Lett. {\bf 99}, 102109 (2011).

\bibitem{sallen}
G. Sallen, L. Bouet, X. Marie, G. Wang, C. R. Zhu, W. P. Han, Y. Lu, P. H. Tan, T. Amand, B. L. Liu, and B. Urbaszek, Phys. Rev. B {\bf 86}, 081301 (2012).

\bibitem{mak3}
K. F. Mak, K. He, C. Lee, G. H. Lee, J. Hone, T. F. Heinz, and J. Shan, Nat. Mater. {\bf 12}, 207 (2013).

\bibitem{he}
K. He, N. Kumar, L. Zhao, Z. Wang, K. F. Mak, H. Zhao, and J. Shan, Phys. Rev. Lett. {\bf 113}, 026803 (2014).

\bibitem{chernikov}
A. Chernikov, T. C. Berkelbach, H. M. Hill, A. Rigosi, Y. Li, O. B. Aslan, D. R. Reichman, M. S. Hybertsen, and T. F. Heinz, Phys. Rev. Lett. {\bf 113}, 076802 (2014).

\bibitem{elliot}
R. J. Elliot, Phys. Rev. {\bf 108}, 1384 (1957).

\bibitem{kulak}
V. D. Kulakovskii, V. G. Lysenk, and Vladislav B. Timofeev, Sov. Phys. Usp. {\bf 28}, 735 (1985).

\bibitem{expT0}
M. Hayne, C. L. Jones, R. Bogaerts, C. Riva, A. Usher, F. M. Peeters, F. Herlach, V. V. Moshchalkov, and M. Henini, Phys. Rev. B {\bf 59}, 2927 (1999).

\bibitem{francoisE}
C. Riva, F. M. Peeters, and K. Varga, Phys. Rev. B {\bf 61}, 13873 (2000); {\it ibid.}, Phys. Status Solidi A {\bf 178}, 513 (2000).

\bibitem{francoisT}
C. Riva, F. M. Peeters, and K. Varga, Phys. Rev. B {\bf 63}, 115302 (2001).

\bibitem{vdwstructure}
A. K. Geim and I. V. Grigorieva, Nature (London) \textbf{499}, 419 (2013).

\bibitem{type1}
J. Kang, S. Tongay, J. Zhou, J. Li, and J. Wu, Appl. Phys. Lett. \textbf{102}, 012111 (2013).

\bibitem{type2}
K. K\'osmider and J. Fern\'andez-Rossier, Phys. Rev. B \textbf{87}, 075451 (2013).

\bibitem{type3}
H. Terrones, F. Lopez-Urias, and M. Terrones, Sci. Rep. \textbf{3}, 1549 (2013).

\bibitem{typeII}
C. Gong, H. Zhang, W. Wang, L. Colombo, R. M. Wallace, and K. Cho, Appl. Phys. Lett. \textbf{103}, 053513 (2013).

\bibitem{type4}
M.-H. Chiu, C. Zhang, H.-W. Shiu, C.-P. Chuu, C.-H. Chen, C.-Y. S. Chang, C.-H. Chen, M.-Y. Chou, C.-K. Shih, and L.-J. Li, Nat. Commun. \textbf{6}, 7666 (2015).

\bibitem{super1}
Y. E .Lozovik, S. L. Ogarkov, and A. A. Sokolik, Phys. Rev. B \textbf{86}, 045429 (2012).

\bibitem{super2}
A. Perali, D. Neilson, and A. R. Hamilton, Phys. Rev. Lett. \textbf{110}, 146803 (2013).

\bibitem{super3}
M. M. Fogler, L. V. Butov, and K. S. Novoselov, Nat. Commun. \textbf{5}, 4555 (2014).

\bibitem{super4}
M. Zarenia, A. Perali, D. Neilson, and F. M. Peeters, Sci. Rep. \textbf{4}, 7319 (2014).

\bibitem{decouple4}
O. L. Berman and R. Y. Kezerashvili, Phys. Rev. B \textbf{93}, 245410 (2016).

\bibitem{exp1}
C.-H. Lee, G.-H. Lee, A. M. Van Der Zande, W. Chen, Y. Li, M. Han, X. Cui, G. Arefe, C. Nuckolls, T. F. Heinz, J. Guo, J. Hone, and P. Kim, Nat. Nanotechnol. \textbf{9}, 676 (2014).

\bibitem{exp2}
M. Palummo, M. Bernardi, and J. C. Grossman, Nano Lett. \textbf{15}, 2794 (2015).

\bibitem{exp3}
P. Rivera, J. R. Schaibley, A. M. Jones, J. S. Ross, S. Wu, G. Aivazian, P. Klement, K. Seyler, G. Clark, N. J. Ghimire, J. Yan, D. G. Mandrus, W. Yao, and X. Xu, Nat. Commun. \textbf{6}, 6242 (2015).

\bibitem{exp4}
H. Heo, J. H. Sung, S. Cha, B.-G. Jang, J.-Y. Kim, G. Jin, D. Lee, J.-H. Ahn, M.-J. Lee, J. H Shim, H. Choi, and M.-H. Jo, Nat. Commun. \textbf{6}, 7372 (2015).

\bibitem{exp5}
A. T. Hanbicki, H.-J. Chuang, M. R. Rosenberger, C. S. Hellberg, S. V. Sivaram, K. M. McCreary, I. I. Mazin, and B. T. Jonker, arXiv: 1802.05310 (2018).

\bibitem{vacuum}
E. Torun, H. P. C. Miranda, A. Molina-S\'anchez, and L. Wirtz, Phys. Rev. B \textbf{97}, 245427 (2018).

\bibitem{sio2}
S. Ovesen, S. Brem, C. Lindera\"alv, M. Kuisma, P. Erhart, M. Selig, and E. Malic, arXiv: 1804.08412v1 (2018).

\bibitem{theory1}
D. Xiao,  G.-B. Liu, W. Feng, X. Xu, and W. Yao, Phys. Rev. Lett. {\bf 108}, 196802 (2012).

\bibitem{decouple1}
J. Sabio, F. Sols, and F. Guinea, Phys. Rev. B \textbf{81}, 045428 (2010).

\bibitem{decouple2}
O. L. Berman, R. Y. Kezerashvili, and K. Ziegler, Phys. Rev. B \textbf{85}, 035418 (2012).

\bibitem{decouple3}
O. L. Berman, R. Y. Kezerashvili, and K. Ziegler, Phys. Rev. A \textbf{87}, 042513 (2013).

\bibitem{previous}
M. Van der Donck, M. Zarenia, and F. M. Peeters, Phys. Rev. B \textbf{96}, 035131 (2017).

\bibitem{screening1}
A. V. Chaplik and M. V. Entin, Zh. Eksp. Teor. Fiz. \textbf{61}, 2496 (1971).

\bibitem{screening2}
L. V. Keldysh, JETP Lett. \textbf{29}, 658 (1979).

\bibitem{screening3}
P. Cudazzo, I. V. Tokatly, and A. Rubio, Phys. Rev. B \textbf{84}, 085406 (2011).

\bibitem{intdist1}
J. He, K. Hummer, and C. Franchini, Phys. Rev. B \textbf{89}, 075409 (2014).

\bibitem{intdist2}
K. Xu, Y. Xu, H. Zhang, B. Peng, H. Shao, G. Ni, J. Li, M. Yao, H. Lu, H. Zhu, C. M. Soukoulis, arXiv: 1804.02518 (2018).

\bibitem{spinsplitv}
Z. Y. Zhu, Y. C. Cheng, and U. Schwingenschl\"ogl, Phys. Rev. B \textbf{84}, 153402 (2011).

\bibitem{berkelbach}
T. C. Berkelbach, M. S. Hybertsen, and D. R. Reichman, Phys. Rev. B {\bf 88}, 045318 (2013).

\bibitem{absorbform}
M. Kira and S. W. Koch, Progress in Quantum Electronics \textbf{30}, 155 (2006).

\bibitem{Eexp}
P. Kim, International Conference on Science and Technology of Synthetic Metals, 2018, Busan, Korea.

\bibitem{trilaag}
D. V. Tuan, M. Yang, and H. Dery, arXiv: 1801.00477 (2018).

\bibitem{transfer}
L. S. R. Cavalcante, A. Chaves, B. Van Duppen, F. M. Peeters, and D. R. Reichman, Phys. Rev. B \textbf{97}, 125427 (2018).

\end{thebibliography}
\end{document}